\documentclass[conference]{IEEEtran}
\IEEEoverridecommandlockouts
\usepackage{cite}
\usepackage{amsmath,amssymb,amsfonts}

\usepackage{algorithmic}
\usepackage{graphicx}
\usepackage{textcomp}
\usepackage{xcolor}
\usepackage{url}
\usepackage{hyperref}
\usepackage{tabularray}
\usepackage{multirow}
\usepackage{comment}
\usepackage{booktabs}
\usepackage{ulem}
\def\BibTeX{{\rm B\kern-.05em{\sc i\kern-.025em b}\kern-.08em
    T\kern-.1667em\lower.7ex\hbox{E}\kern-.125emX}}

\begin{document}

\title{TrueFake: A Real World Case Dataset of Last Generation Fake Images also Shared on Social Networks}
\author{
\IEEEauthorblockN{Stefano Dell'Anna}
\IEEEauthorblockA{
\textit{University of Trento}\\
Trento, Italy \\
stefano.dellanna@unitn.it}
\and
\IEEEauthorblockN{Andrea Montibeller}
\IEEEauthorblockA{
\textit{University of Trento}\\
Trento, Italy \\
andrea.montibeller@unitn.it}
\and
\IEEEauthorblockN{Giulia Boato}
\IEEEauthorblockA{
\textit{University of Trento and Truebees srl}\\
Trento, Italy \\
giulia@truebees.eu}
}

\maketitle

\begin{abstract}
 AI-generated synthetic media are increasingly used in real-world scenarios, often with the purpose of spreading misinformation and propaganda through social media platforms, where compression and other processing can degrade fake detection cues. Currently, many forensic tools fail to account for these in-the-wild challenges. In this work, we introduce TrueFake, a large-scale benchmarking dataset of 600,000 images including top notch generative techniques and sharing  via three different social networks. This dataset allows for rigorous evaluation of state-of-the-art fake image detectors under very realistic and challenging conditions. Through extensive experimentation, we analyze how social media sharing impacts detection performance, and identify current most effective detection and training strategies. Our findings highlight the need for evaluating forensic models in conditions that mirror real-world use.

\end{abstract}


\begin{figure}[t]
  \includegraphics[width=\linewidth]{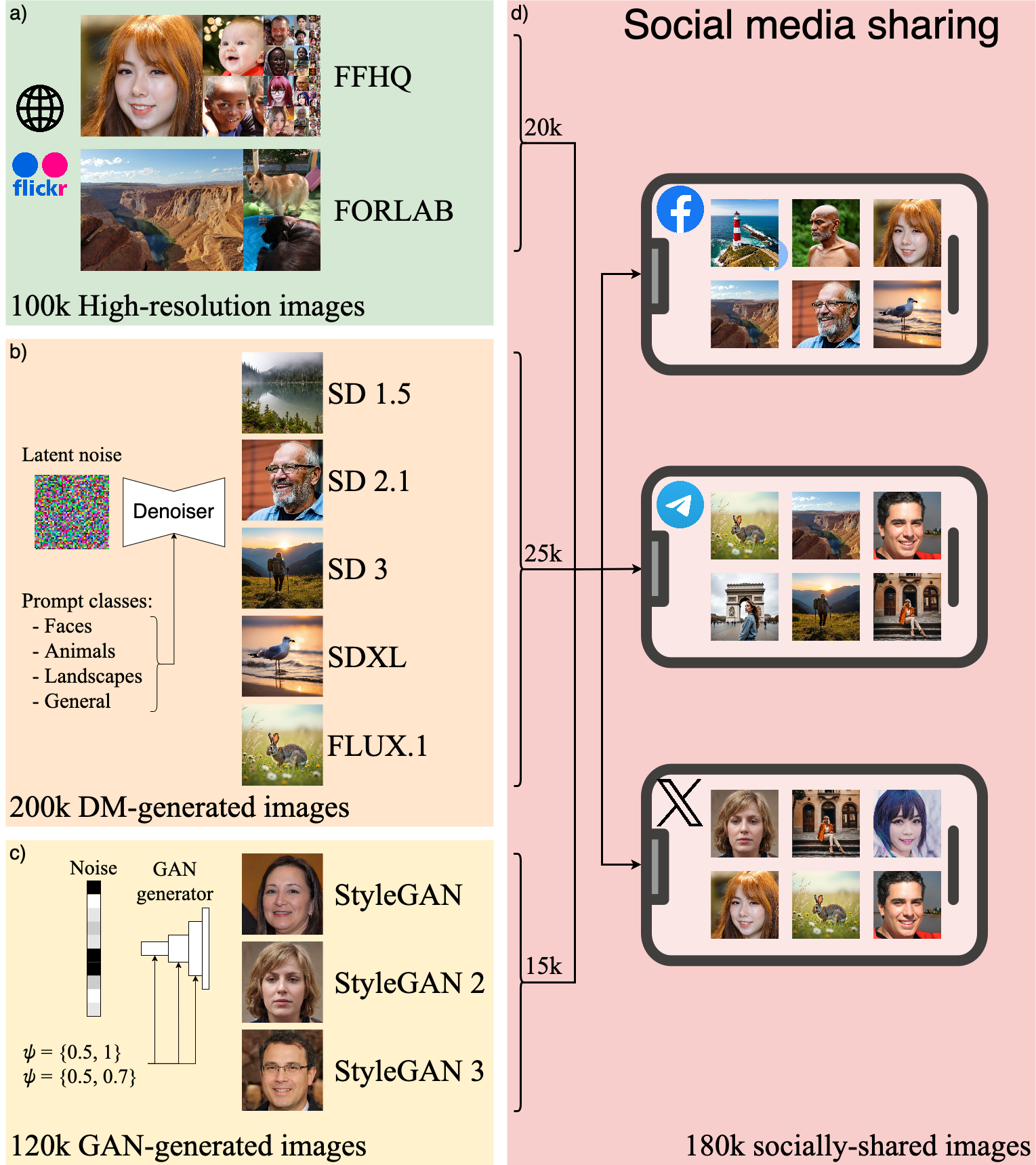}
  \caption{The \textit{TrueFake} dataset contains a total of 600,000 images: a) 100,000 real images sourced from the FORLAB \cite{iuliani2021leak} and FFHQ \cite{ffhq-dataset} datasets, b) 200,000 images generated using five DM-based techniques (Stable Diffusion (versions 1.5 to 3), Stable Diffusion XL, and FLUX.1), c) 120,000 images produced with GANs (StyleGAN 1, 2, and 3), and d) 180,000 of them shared via social media platforms. Indeed, a subset of 20,000 real, 25,000 DM-generated, and 15,000 GAN-generated images were also shared on three social networks (Facebook, X, and Telegram) by using the respective APIs. 
}
  \label{fig1:visual_abstract}
\end{figure}

\section{Introduction}
\label{sec:intro}

In recent years, AI-generated media (such as images, videos, and audio) have increasingly become part of everyday life \cite{fabuyi2024deepfake} becoming widely used in the entertainment industry, including movie production and advertising.

The literature provides a broad range of AI media generators capable of producing hyper-realistic images \cite{karras_style-based_2019,rombach_high-resolution_2022}, videos \cite{brooks_video_2024}, and even audio \cite{audio_DM}. AI image generation is typically achieved using Generative Adversarial Networks (GANs) \cite{karras_style-based_2019}, which synthesize images from random noise, or Diffusion Models (DMs) \cite{rombach_high-resolution_2022}, that generate highly realistic images from textual descriptions. Recently, DM-based generative techniques have also been extended to video and audio synthesis, achieving remarkable levels of realism \cite{lago2021more,nightingale2022ai}.

The widespread availability of GAN- and DM-based generative techniques has made it possible for anyone to create various forms of AI-generated media. While this accessibility fosters creativity and the production of high-quality content, it also introduces significant challenges. These technologies are frequently exploited to produce fake media for misleading propaganda, hate speech dissemination, identity impersonation, and fraudulent activities \cite{amerini2024deepfake, amerini2025deepfake}. Moreover, recent studies illustrates how humans are poorly equipped to distinguish any kind of AI-generated media \cite{barrington2024people,lago2021more,nightingale2022ai}. To mitigate the misuse of AI media generators, researchers have proposed numerous AI-based detection methods that analyze either low-level features or high-level semantic features of media content (see, for instance, \cite{gragnaniello_are_2021,cozzolino_raising_2024}). 

However, existing detection techniques often operate under controlled laboratory conditions and fail to fully account for real-world scenarios. In practice, AI-generated media are frequently shared via social networks, which apply proprietary processing and compression techniques to optimize storage allocation \cite{10007988}. Nevertheless, studies have shown that these social network media processing significantly degrade the accuracy of fake media detectors by obscuring the features they rely on \cite{10007988}.

Motivated by these observations, we introduce a novel large-scale benchmarking dataset called \textit{TrueFake}, depicted in Fig.~\ref{fig1:visual_abstract}, comprising {600,000} images, including those shared on three social networks. This dataset enables the evaluation of state-of-the-art (SoA) fake image detection methods under real-world and adversarial conditions. {Indeed, TrueFake includes high-quality real images, also processed by AI-based image enhancement firmware \cite{ffhq-dataset,iuliani2021leak}, and cutting-edge GAN- and DM-generated images} \cite{karras_style-based_2019,karras_analyzing_2020,karras_alias-free_2021,rombach_high-resolution_2022,podell2023sdxl,esser2024scaling,flux2024}, in both cases also shared on social networks. In details, we distributed a subset of {180,000} images on three social media platforms: Telegram, Facebook, and X (formerly Twitter).

In our experiments, we analyze the performance degradation of several SoA fake image detectors, trained on uncompressed datasets, when tested on social media compressed images. Thus, we provide insights into the most effective detection and training pipelines and highlight the importance of evaluating {forensic detectors} using datasets that reflect real-world scenarios.

The remainder of this paper is structured as follows: Section~\ref{sec:soa} reviews related works, while Section~\ref{sec:dataset} introduces the proposed dataset. Section~\ref{sec:detection_methods} describes the fake image detection methods {under test, while Section~\ref{sec:experimental_results} shows their performance on TrueFake}. Finally, we conclude with a discussion in Section~\ref{sec:conclusions}.



\section{Related works}
\label{sec:soa}
Detecting synthetic multimedia shared on social networks presents a significant real-world challenge~\cite{amerini2024deepfake, amerini2025deepfake}. {Indeed, }to manage bandwidth and storage constraints, these platforms employ aggressive compression and data processing techniques~\cite{10007988}. While such techniques effectively improve storage allocation, they also degrade forensic features crucial for distinguishing real content from manipulated or AI-generated media~\cite{10007988, Verdoliva2020910, maier2024reliable}.  

One of the earliest studies on the impact of social media compression on GAN-generated images~\cite{8397040} highlighted the detrimental effects of these processes on forensic detection methods. The study found that while visual quality remained largely unaffected, forensic traces essential for fake image detection were significantly diminished. These findings were further explored in~\cite{10007988}, where the authors compiled a dataset {of only faces} consisting of 80,000 fake images generated using StyleGAN models and 70,000 real images sourced from SoA datasets~\cite{karras_style-based_2019}. The study also examined the extent of compression applied by platforms such as Twitter, Facebook, and Telegram, demonstrating that fine-tuning detection models on their dataset helped preserve detection accuracy despite these transformations. Notably, the authors emphasized the importance of mitigating \textit{catastrophic forgetting loss}~\cite{kirkpatrick2017overcoming} when adapting detectors to evaluate social network processed images. Nonetheless, the {limited applicability of \cite{10007988} (as it focuses solely on facial images)}, along with the continuous evolution {of image generation techniques and} social network image processing pipelines, has quickly rendered this dataset outdated.

Recently, the proliferation of AI-generated fake images on social networks has raised concerns about their misuse in impersonation and inauthentic online activities~\cite{Yang_Singh_Menczer_2024}. This has fueled growing research interest in detection methodologies. Yang et al.~\cite{Yang_Singh_Menczer_2024} conducted a systematic analysis of fake accounts using GAN-generated faces on X (formerly Twitter). They proposed {an} heuristic approach based on the eye alignment patterns found in GAN-generated faces and estimated that approximately 10,000 such accounts were active daily on X. However, more recent GAN architectures, such as StyleGAN3, have mitigated the vulnerabilities targeted by this heuristic.  
Similarly, Ricker et al.~\cite{ricker2024ai} performed a large-scale study of AI-generated profile images on X, proposing a multi-stage detection pipeline using a ResNet50~\cite{koonce2021resnet}. Their study produced estimates of accounts utilizing GAN-generated profile pictures that {are} closely aligned with the findings of Yang et al.~\cite{Yang_Singh_Menczer_2024}.  

Building upon these efforts, Maier and Riess~\cite{maier2024reliable} expanded the scope of fake image detection beyond GAN-generated content to include images synthesized using diffusion models. They introduced a Bayesian Neural Network for detecting synthetic images, demonstrating performance comparable to that of the ResNet50 model used in~\cite{ricker2024ai}.  
Additionally, Kumar et al.~\cite{kumar2023gan} explored deepfake generation techniques and their detection using GAN-based deep convolutional models. {Their study conducted a comparative analysis of different models, providing valuable insights into their relative strengths and weaknesses in detection performance.}

Despite significant advancements in synthetic image detection on social networks, the problem remains far from solved. Social media platforms frequently update their image processing algorithms, necessitating continuous adaptation and fine-tuning of detection models~\cite{10007988}. Moreover, many existing detectors have been tested predominantly on GAN-generated images, leaving their generalization capabilities on DM-generated images uncertain. Furthermore, the rapid development of new generative models, often capable of producing increasingly photorealistic images, underscores the urgent need for highly generalizable detection systems.  

To address these challenges, we introduce \textit{TrueFake}, a novel, large-scale, and diverse dataset of real and AI-generated images shared on social media platforms. This dataset serves as a benchmark for evaluating the robustness of SoA detectors across a wide range of generative techniques. The following sections provide a detailed overview of TrueFake and its potential applications in advancing forensic detection methods.

\section{Dataset}
\label{sec:dataset}



The TrueFake dataset\footnote{The dataset {is available at \url{https://github.com/MMLab-unitn/TrueFake-IJCNN25}}} (see Fig.~\ref{fig1:visual_abstract}) consists in 600,000 images also shared on social networks. In detail, 100,000 real images were sourced from the FFHQ \cite{ffhq-dataset} and FORLAB \cite{iuliani2021leak} datasets, covering diverse scenes, resolutions, and camera acquisition pipelines. Additionally, 200,000 images depicting faces, animals, landscapes, and everyday life scenes were generated using five SoA DM-based techniques, while 120,000 face-only images were produced by three GAN-based image generators. Finally, a subset of 60,000 of these real and fake images was shared across three social networks (Facebook, X, and Telegram) via their respective APIs, resulting in a total of 180,000 shared images.

To the best of our knowledge, TrueFake is one of the few publicly available and up-to-date datasets that includes a large and diverse variety of image scenes (e.g. faces, animals, landscapes, and everyday life), enabling the evaluation of {forensic} detectors on both advanced generative models and social network-processed images. Additional details on TrueFake images are provided in Table~\ref{tab:dataset_details} {and in the following sub-sections.} 



\subsection{Real Images Collection}
TrueFake's real images were collected from the FFHQ \cite{ffhq-dataset} and FORLAB \cite{iuliani2021leak} datasets to ensure a diverse representation in terms of scene and camera acquisition processing. From FFHQ, which contains high-quality facial images from Flickr, we selected 70,000 images, while from FORLAB, we selected 30,000 images representing various scenes and resolutions, captured by different mobile and DSLR cameras, and subjected to both AI-based and in-camera and out-camera post-processing enhancements \cite{montibeller2024shedding}. These post-processing enhancements help better model the feature space of real images captured by modern devices.

\subsection{GAN-generated Images}
\label{sec:gans}
The first three selected generative techniques are based on GANs \cite{goodfellow_generative_2020}, a class of neural networks capable of generating images from noise. We specifically used StyleGAN (SG) \cite{karras_style-based_2019}, StyleGAN2 (SG2) \cite{karras_analyzing_2020}, and StyleGAN3 (SG3) \cite{karras_alias-free_2021}, all trained on FFHQ, to exclusively generate hyper-realistic facial images. Moreover, the inclusion of all three versions allows us to evaluate the impact of architectural improvements across different StyleGAN generations. Indeed, StyleGAN2 mitigates the ``texture sticking" artifacts of StyleGAN by replacing the Adaptive Instance Normalization with weight demodulation and removing progressive growing. Similarly, StyleGAN3 addresses StyleGAN2 aliasing issues by ensuring geometric transformation equivariance through enhanced up/downsampling and Fourier features.

During the GAN images generative process, each StyleGAN model generated 40,000 images at a resolution of $1024 \times 1024$ pixels. The 40,000 images were divided in two subsets of 20,000 images, each generated using two different values of $\psi$, which control the trade-off between image quality and diversity. $\psi$ was set equals to $\{0.5, 0.7\}$ for SG and SG3, and $\{0.5, 1\}$ for SG2.

\subsection{DM-generated Images}
The remaining synthetic images were generated using five DMs able to generate hyper-realistic images from textual prompts: Stable Diffusion 1.5 (SD1.5) \cite{rombach_high-resolution_2022}, Stable Diffusion 2.1 (SD2.1) \cite{rombach_high-resolution_2022}, Stable Diffusion 3 (SD3) \cite{esser2024scaling}, Stable Diffusion XL (SDXL) \cite{podell2023sdxl}, and FLUX.1 \cite{flux2024}. As {underlined in the previous subsection}, including multiple versions of Stable Diffusion enables us to analyze how updates in the generator architecture impact the performance of fake image detectors. Stable Diffusion updates with respect to precedent versions include better image coherence in SD2.1, enhanced photorealism via a two-stage pipeline in SDXL, and superior text-image alignment through multimodal diffusion transformers in SD3. {On the other hand}, FLUX.1 was selected for its ability to generate high-quality, diverse images with strong prompt adherence, leveraging rotary positional embeddings and parallel attention layers.

For the generation of fake DM images, we used 40,000 textual prompts \footnote{Code for prompt generation {is available at \url{https://github.com/MMLab-unitn/TrueFake-IJCNN25}}} equally divided in four categories: \textit{Animals}, \textit{Faces}, \textit{Landscapes}, and \textit{General}. Prompts for \textit{Animals} were obtained using an algorithm combining databases of habitat-species pairings and ecological constraints, while \textit{Faces} prompts taking databases of structured demographic and aesthetic variations. \textit{Landscapes} prompts combine various environmental-weather-temporal attributes, in contrast, \textit{General} prompts were sourced from a public image-description dataset \footnote{\url{https://huggingface.co/datasets/sezenkarakus/image-description-dataset-v2}}.


During image generation, all models were instantiated in \textit{float16}, except for FLUX.1, which used \textit{bfloat16}. 
The number of inference steps for each DMs are: 100 for SD1.5 and SD2.1; 28 for SD3 (medium version); 50 for SDXL (base-1.0); and 28 for FLUX.1 (dev version). The resolution of the generated images is reported in Table \ref{tab:dataset_details}.

\subsection{Social Network Sharing}
\label{sec:dataset_shared}

A subset of 60,000 images, sampled from both real and synthetic ones, was shared on three social networks (totaling 180,000 shared images) to analyze the impact of social network image processing and compression on fake image detection algorithms. For each generative technique, 5,000 images equally distributed among the four prompt categories (i.e. \textit{Animals}, \textit{Landscapes}, \textit{Faces}, and \textit{General}) and $\psi$ values, were uploaded and downloaded from Facebook, X, and  Telegram via their respective APIs. Additionally, 20,000 real images (equally splitted between FORLAB and FFHQ) were also shared across these platforms.

To quantify the impact of social network image processing and compression, in Table~\ref{tab:dataset_social}, we present key metrics such as average Peak Signal-to-Noise Ratio (PSNR), Mean Square Error (MSE), and Structural Similarity Index (SSIM) \cite{hore2010image} estimated using images before and after social media compression. Table~\ref{tab:dataset_social} also includes JPEG Quality Factor (QF) estimates. Interestingly, while X and Telegram use consistent JPEG QF values for compression, Facebook dynamically adjusts the QF based on input image characteristics also resulting in worst results in terms of PSNR, MSE, and SSIM values.

Finally, we observed that each social network apply resizing under specific conditions. Given an input image of resolution $W \times H$, where $W$ and $H$ represent its width and height, each social network {outputs} an image of size $w\times h$. However, Facebook and X, if $W > w$ with $w$ set to 720 pixels for Facebook and 1200 pixels for X, {downscale} the input image such that $h=H\cdot \alpha$, with downscaling parameter $\alpha = w/W$. In contrast, Telegram downscales the input image such that $h=800$ if $H>h$, and $w=W\cdot \beta $ with downscaling parameter $\beta = h/H$. {{Notice that,} X does not apply any image resizing or compression when images have resolution $\leq768 \times 768$ pixels.}


\begin{table}[t]
\caption{TrueFake Dataset Overview. ``Source" refers to the origin of real images or the fake image generation method. ``Input" specifies key parameters used for generating fake images, while "Resolution" indicates images resolutions. The ``Non-Shared" and ``Shared" columns summarize the number of images respectively not-shared and shared on each social network.}
\centering
\begin{tblr}{rows = {abovesep=2pt,belowsep=1pt}, colsep=1pt, colspec={X[0.8,l,m]X[1.1,l,m]X[1,c,m]X[1,c,m]X[1,c,m]}}
Source & Input & Resolution & Non-Shared & Shared \\ 
\hline \hline
FFHQ & & $1024^2$ & 70k & 3$\times$10k \\ 
\hline 
FORLAB & & {variable} & 30k & 3$\times$10k \\ 
\hline\hline 
{SG \\ SG2 \\ SG3} & {$\psi=\{0.5, 0.7\}$ \\ $\psi=\{0.5, 1\}$ \\ $\psi=\{0.5, 0.7\}$} & { $1024^2$  }  & {120k} & {3$\times$3$\times$5k} \\ 
\hline
{SD 1.5 \\ SD 2.1 \\ SD 3 \\ SD XL \\ FLUX} & {Prompts class: \\ - Face \\ - Animal \\ - Landscape \\ - General} & {$512^2$ \\ $768^2$ \\ $1024^2$ \\ $1024^2$ \\ $1024^2$} & {200k} & {3$\times$5$\times$5k} \\ \hline\hline
Total & & & {100k real \\ 320k fake} & {60k real \\ 120k fake}
\end{tblr}
\label{tab:dataset_details}
\end{table}

\begin{table}[t]
\caption{Compression metrics for socially shared images. The first three values (PSNR, MSE, and SSIM) represent average metrics, while JPEG QF indicates ranges of quality factors used by each social networks.}
\centering
\begin{tblr}{rows = {abovesep=2pt,belowsep=1pt}, colspec={X[1,l,m]X[1,c,m]X[1,c,m]X[1,c,m]X[1,c,m]}}
Social      & PSNR  & MSE   & SSIM & JPEG QF                \\ 
\hline \hline
Facebook    & 36.21 & 17.63 & 0.92 & \{{61} - 92\}   \\ 
\hline
X           & 39.26 & 9,29  & 0.95 & {87}            \\ 
\hline
Telegram    & 37.09 & 15.46 & 0.93 & {85}            \\ 
\hline\hline
\end{tblr}
\label{tab:dataset_social}
\end{table}

\section{Detection methods}
\label{sec:detection_methods}
To evaluate the respective performance degradation of fake image detectors trained on laboratory conditions and tested in real-world scenarios, including social network image processing and compression, we selected five methods representative of the current {forensic} SoA. 
The first three methods exploit low-level features extracted by CNN-based architectures like ResNet50 (R50-ND) 
\cite{corvi_detection_2023}, MISLNet \cite{bayar_constrained_2018}, and Neighboring Pixel Relationships (NPR) \cite{tan_rethinking_2024}. The other two very recent methodologies, CLIP-based detector (CLIP-D) \cite{cozzolino_raising_2024} and Prompt2Guard (P2G) \cite{laiti_conditioned_2025}, exploit high-level semantic features retrieved using CLIP \cite{clip} Vision Transformer (ViT) to classify real from fake images. Finally, inspired by recent findings of \cite{cozzolino_raising_2024}, we propose R50-E2P, a detector that leverages a freezed ResNet50 pretrained on ImageNet \cite{imagenet15russakovsky} as a feature encoder for fake image detection.
{Next subsections describe novelty and architectures of each above mentioned and tested approach.}\\

\subsubsection{R50-ND}
In \cite{corvi_detection_2023}, the authors demonstrate that significant improvements in accuracy and generalization to unseen generative image models can be achieved by modifying the standard ResNet50 architecture \cite{he_deep_2016} and employing extensive data augmentation techniques. Specifically, the approach proposed in \cite{corvi_detection_2023} eliminates the downsampling operation in the first layer of the original ResNet50 architecture and incorporates various data augmentation techniques, including JPEG compression, Gaussian blur, Gaussian noise, random cutout and color jitter.
By removing the downsampling operation in the first layer, the authors show that low-level traces in synthetic images, which are crucial for fake image detection, are better preserved. Consequently, extensive data augmentation enhances the detection capabilities of the proposed method, particularly in challenging scenarios involving strong JPEG compression and resizing.\\

\subsubsection{MISLNet}

Very recently, MISLNet \cite{vahdati_beyond_2024}, a network initially developed for image manipulation detection \cite{bayar_constrained_2018} and camera source attribution \cite{Fang_2023_BMVC}, has been applied to fake video detection, demonstrating outstanding performance. 
This success is attributed to the integration of the so-called constrained convolutional layer within the MISLNet framework. These layers provide high adaptability to a variety of tasks (including source attribution \cite{bayar_constrained_2018}, tampering detection \cite{Fang_2023_BMVC}, and fake media detection \cite{vahdati_beyond_2024}) by acting as learned prediction error filters which function similarly to low-level forensic traces \cite{bayar_constrained_2018}. 
Intrigued by these properties and the promising results obtained in fake video detection (where detectors are often challenged by inter-frame dependencies and compression artifacts), we incorporated MISLNet into our experiments, by re-training it on the TrueFake dataset.\\

\subsubsection{NPR}

In \cite{tan_rethinking_2024}, the authors propose leveraging up-sampling artifacts introduced by image generative models. They demonstrate that up-sampling operations during image generation induce local pixel interdependencies in synthetic images, which NPR captures through spatial differences within small image patches (i.e., $2\times2$ grids). This, in turn, allows the classification network (a standard ResNet-50) to better focus on the local correlations between pixels. 
Unlike prior frequency-based methods, NPR emphasizes localized structural artifacts, enabling robust detection capabilities across diverse generative models. However, the original paper \cite{tan_rethinking_2024} does not mention any tests involving compressed or processed images. {This makes testing the technique on our dataset particularly relevant, as it allows us to evaluate its performance under practical real-world conditions.  }\\



\subsubsection{CLIP-D}
Recently, Cozzolino et al. \cite{cozzolino_raising_2024} proposed to use the CLIP \cite{clip} pre-trained ViT as high-level semantic feature extractor and train a shallow network upon composed by a single linear layer to classify real from fake images. In their paper, the authors demonstrated remarkable generalization capabilities to unknown generative techniques as well as robustness to image processing and few-shot abilities.\\


\subsubsection{P2G}
Drawing inspiration from \cite{cozzolino_raising_2024}, Prompt2Guard (P2G) \cite{laiti_conditioned_2025} is a method developed for continual fake image detection, leveraging the CLIP Vision-Language Model (VLM) \cite{clip}.  In their framework \cite{laiti_conditioned_2025}, the authors employed conditioned textual prompts in the form of ``a \{real/fake\} photo of a \{$c*$\}", where $c*$ represents the set of the five most similar CLIP-estimated image categories from ImageNet-1k \cite{imagenet15russakovsky}. The VLM then utilizes these conditional prompts, in combination with the input image, to focus on features that better distinguish real from synthetic images.  
This approach is enabled by a training process in which P2G learns to maximize the similarity between the visual embeddings of the input image and the corresponding textual embeddings, extracted by the vision and text encoders, respectively.  
Similar to NPR, the proposed solution demonstrated promising results in terms of efficiency, accuracy, and scalability when tested on uncompressed generated images, making it a compelling approach for continual fake image detection.\\




\subsubsection{ResNet50 as Feature Encoder to Prototypes (R50-E2P)}
\label{sec:our_method}
Building on the recent strategy that effectively leverages ViT \cite{cozzolino_raising_2024}, this study explores whether similar improvements in generalization and robustness to unseen generators and data processing can be achieved using R50-ND with frozen ImageNet-pretrained weights as a feature encoder.

{Indeed, \cite{cozzolino_raising_2024} demonstrates that ViT-based forensic detectors, pre-trained on large datasets of non-synthetic media, inherently encode fundamental properties of real images, and these properties not only help distinguish real from fake images but also enable superior generalization to unseen generative pipelines compared to fully retrained detectors like R50-ND.}
However, since R50-ND was also pre-trained on ImageNet \cite{imagenet15russakovsky} (another large dataset of real images) we hypothesize that its weaker generalization ability is primarily due to catastrophic forgetting \cite{kirkpatrick2017overcoming}.

Specifically, features extracted by R50-ND will be classified by a shallow, trainable classification head consisting of a single fully connected layer. This layer reduces the feature dimensionality from 2048 to 128 while implementing learned prototypes \cite{lee2020multi,ruff2018deep}. Learned prototypes are commonly used in deep learning classification tasks to detect out-of-distribution samples by modeling an isotropic Gaussian class-conditional distribution representative of the input data.


If \(\mathbf{f}\) represents the features extracted by the fully connected layer of R50-ND with frozen weights, the corresponding learned prototypes are computed as follows:
\begin{equation}  
s(\mathbf{f}) = \left[ \operatorname{sgn} \left( \sum_{k=1}^{D} (f_k - c_k) \right) \cdot \left( \frac{\|\mathbf{f} - \mathbf{c}\|_2}{2\sigma_k^2} \right) + D \ln \sigma_k \right]  
\label{eq:proto_new}  
\nonumber
\end{equation}  
where, \(D\) denotes the input feature dimension, while \(\mathbf{c} = [c_1, \dots, c_D] \in \mathbb{R}^{D}\) represents a set of centroids characterizing an isotropic Gaussian class-conditional distribution with standard deviations \(\boldsymbol{\sigma} = [\sigma_1, \dots, \sigma_D] \in \mathbb{R}^{D}\).
Both \(\mathbf{c}\) and \(\boldsymbol{\sigma}\) are learned during training and are constrained by \(\log \sigma_k > 0, \forall k\), ensuring that the distance function satisfies the triangle inequality, an essential property for using learned prototypes as a valid metric \cite{lee2020multi}. Finally, we incorporated the term \(\operatorname{sgn} \left( \sum_{k=1}^{D} (f_k - c_k) \right)\) to enhance performance in binary classification tasks \footnote{Full training and testing code {is available at \url{https://github.com/MMLab-unitn/TrueFake-IJCNN25}}}.




\section{Experimental results}
\label{sec:experimental_results}

In this section, {we evaluate the performance in terms of {single-class True Positive Rate (for fake images) and True Negative Rate (for real images),} of all fake image detectors described in Sect.~\ref{sec:detection_methods}.} All {methods} are trained on non-shared images (under ideal laboratory conditions) and tested first on non-shared images, then on images shared across three social networks: Facebook, X, and Telegram. Our objective is to assess the impact of social {media sharing} on the ability of SoA detectors to differentiate between real and fake images.

To isolate the effect of social network image processing and ensure that any decrease in {TPR and TNR} is not due to poor generalization when trained only on GAN- or DM-generated images, all detectors were trained on fake images from StyleGAN2 and Stable Diffusion XL, as well as real images from FFHQ and FORLAB.
The training, validation, and test sets followed a 70/15/15 split, with images that were also shared on social networks used only during testing.

{Detectors were trained using the code and pipeline provided in their respective papers, with the only exception being MISLNet, for which we applied the same data augmentation as the here proposed R50-E2P, thus} including random scaling in the range {0.5–1.5}, random cropping to (300, 300), color jitter, Gaussian blur, and JPEG compression.
 All experiments were conducted on a server with the following specifications: NVIDIA GeForce RTX 4090, a 24-core Intel(R) Core(TM) i9-14900KF CPU @ 6.0GHz, and 128 GB RAM.

\subsection{Results on non-shared images}
\label{sec:non-shared}
\begin{table}[t]
\caption{{TPR and TNR} performances on non-shared images for detectors trained using images from FFHQ and FORLAB, and from SG2 and SDXL. Best results in \textbf{bold}}
\begin{tblr}{rows = {abovesep=2pt,belowsep=1pt}, colsep = 1pt, colspec={X[1.5,l,m]X[1,c,m]X[1,c,m]X[1,c,m]X[1,c,m]X[1,c,m]X[1,c,m]}}
Data        & {R50-E2P} & R50-ND         & MISLNet       & NPR           & CLIP-D        & P2G           \\ \hline \hline 
FFHQ        & 0.96 & \textbf{1.00}  & 0.99          & \textbf{1.00} & 0.95          & 0.99          \\ 
FORLAB      & 0.99 & \textbf{1.00}  & 0.97          & 0.98          & 0.98          & \textbf{1.00} \\ \hline 
Average TNR & 0.98 & \textbf{1.00}  & 0.98          & 0.99          & 0.97          & \textbf{1.00} \\ \hline \hline
SG          & 0.98 & 0.94           & 0.59          & 0.54          & \textbf{0.99} & 0.92          \\ 
SG2         & 0.99 & \textbf{1.00}  & \textbf{1.00} & \textbf{1.00} & 0.94          & 0.93          \\ 
SG3         & 0.59 & 0.00           & 0.00          & 0.02          & \textbf{0.96} & 0.82          \\ 
SD1.5       & 0.91 & \textbf{0.99}  & 0.72          & 0.43          & 0.88          & 0.88          \\ 
SD2.1       & 0.78 & 0.78           & 0.37          & 0.41          & 0.85          & \textbf{0.90} \\ 
SD3         & 0.63 & 0.64           & 0.32          & 0.55          & 0.84          & \textbf{0.87} \\ 
SDXL        & 0.99 & \textbf{1.00}  & 0.99          & \textbf{1.00} & \textbf{1.00} & \textbf{1.00} \\ 
FLUX.1      & 0.92 & 0.94           & 0.86          & 0.86          & \textbf{0.98} & 0.93          \\ \hline 
Average TPR & 0.85 & 0.79           & 0.61          & 0.60          & \textbf{0.93} & 0.91          \\ \hline \hline

\end{tblr}
\label{tab:results_pristine}
\end{table}

In Table \ref{tab:results_pristine}, we report the {TPR and TNR} achieved by each detector when trained and tested on non-shared images, reflecting the so-called ideal laboratory conditions.
We observe that all detectors achieve {TPR and TNR} values $\geq 0.93$ when tested on image classes seen during training (e.g. FFHQ, FORLAB, SG2, SDXL). In contrast, when evaluating unseen generators, ViT-based methods (CLIP-D and P2G) demonstrate superior generalization capabilities, achieving {positive and negative rate} values comparable to those obtained for generators seen during training.

Focusing on CNN-based detectors ({R50-E2P}, R50-ND, MISLNet, and NPR), we observe that {R50-E2P} and R50-ND achieve the highest average {TPR and TNR}. This aligns with findings from \cite{gragnaniello_are_2021}, which highlight the benefits of starting from CNN pretrained on large dataset of real data like ImageNet \cite{imagenet15russakovsky}.  In contrast, MISLNet, being completely trained from scratch, performs worse while, the lower performance of NPR can be attributed to {the domain applicative shift, which occurs when re-training a ResNet50 (pretrained for ImageNet classification) to identify neighboring pixel relationships, and was mitigated by} the significantly larger and more semantically diverse training dataset used in the original paper \cite{tan_rethinking_2024}.

Morevoer, examining the results on SG3, {R50-E2P} is the only CNN-based detector capable of effectively identifying these images. As discussed in Sect.~\ref{sec:gans}, the performance on StyleGAN3 are influenced by architectural modifications aimed at minimizing artifacts commonly found in earlier versions of StyleGAN images \cite{karras_alias-free_2021}. While training R50-ND on a larger and more diverse set of GAN-generated images, as proposed in \cite{corvi_detection_2023}, could mitigate StyleGAN3 detection challenges, {R50-E2P} offers a simpler yet effective few-shot solution, similarly to CLIP-D \cite{cozzolino_raising_2024} where a pretrained network is used as feature encoder.

Ultimately, while overall performance remains lower than CLIP-D, we believe the results obtained by {R50-E2P} could open new avenues for leveraging any pretrained networks as feature encoder in fake image detection tasks.

\subsection{Results on shared images}

\begin{table}[t]
\caption{{TPR and TNR obtained} on images shared via social networks, with deltas (shown as subscripts) indicating changes relative to non-shared images. All detectors were trained on non-shared images from FFHQ, FORLAB, SG2, and SDXL. When Deltas absolute values exceeds 0.05, positive ones are highlighted in \textcolor{blue}{blue}, while negative deltas in \textcolor{red}{red}.}
\begin{tblr}{rows = {abovesep=1pt,belowsep=1pt}, colsep = 2pt, colspec = {X[1,l,m]X[-1,l,m]X[1,c,m]X[1,c,m]X[1,c,m]X[1,c,m]|X[1,c,m]X[1,c,m]}}
\SetCell[c=2]{c} {\footnotesize Data \\\tiny{Metric}} & & {R50-E2P} & R50-ND & MISLNet & NPR & CLIP-D & P2G \\ \hline \hline 
\SetCell[r=3]{r} {\footnotesize FFHQ \\\tiny{TNR}} & \tiny FB & \scriptsize \textcolor{red}{0.74 \tiny -0.22} & \scriptsize \textcolor{black}{0.96 \tiny -0.04} & \scriptsize \textcolor{red}{0.41 \tiny -0.58} & \scriptsize \textcolor{black}{1.00 \tiny +0.00} & \scriptsize \textcolor{red}{0.79 \tiny -0.16} & \scriptsize \textcolor{black}{1.00 \tiny +0.01} \\ 
 & \tiny TL & \scriptsize \textcolor{black}{0.95 \tiny -0.01} & \scriptsize \textcolor{black}{1.00 \tiny +0.00} & \scriptsize \textcolor{red}{0.79 \tiny -0.20} & \scriptsize \textcolor{black}{1.00 \tiny +0.00} & \scriptsize \textcolor{black}{0.93 \tiny -0.02} & \scriptsize \textcolor{black}{1.00 \tiny +0.01} \\ 
 & \tiny X & \scriptsize \textcolor{black}{0.98 \tiny +0.02} & \scriptsize \textcolor{black}{1.00 \tiny +0.00} & \scriptsize \textcolor{red}{0.70 \tiny -0.29} & \scriptsize \textcolor{black}{1.00 \tiny +0.00} & \scriptsize \textcolor{black}{0.93 \tiny -0.02} & \scriptsize \textcolor{black}{1.00 \tiny +0.01} \\ \hline 
\SetCell[r=3]{r} {\footnotesize FORLAB \\ \tiny{TNR}} & \tiny FB & \scriptsize \textcolor{red}{0.90 \tiny -0.09} & \scriptsize \textcolor{red}{0.89 \tiny -0.11} & \scriptsize \textcolor{red}{0.43 \tiny -0.54} & \scriptsize \textcolor{black}{1.00 \tiny +0.02} & \scriptsize \textcolor{black}{0.99 \tiny +0.01} & \scriptsize \textcolor{black}{1.00 \tiny +0.00} \\ 
 & \tiny TL & \scriptsize \textcolor{red}{0.93 \tiny -0.06} & \scriptsize \textcolor{black}{0.97 \tiny -0.03} & \scriptsize \textcolor{red}{0.55 \tiny -0.42} & \scriptsize \textcolor{black}{1.00 \tiny +0.02} & \scriptsize \textcolor{black}{0.99 \tiny +0.01} & \scriptsize \textcolor{black}{1.00 \tiny +0.00} \\ 
 & \tiny X & \scriptsize \textcolor{black}{0.98 \tiny -0.01} & \scriptsize \textcolor{black}{0.98 \tiny -0.02} & \scriptsize \textcolor{red}{0.78 \tiny -0.19} & \scriptsize \textcolor{black}{1.00 \tiny +0.02} & \scriptsize \textcolor{black}{0.99 \tiny +0.01} & \scriptsize \textcolor{black}{1.00 \tiny +0.00} \\ \hline \hline 
\SetCell[r=3]{r} {\footnotesize SG \\ \tiny{TPR}} & \tiny FB & \scriptsize \textcolor{red}{0.69 \tiny -0.29} & \scriptsize \textcolor{black}{0.96 \tiny +0.02} & \scriptsize \textcolor{blue}{0.79 \tiny +0.20} & \scriptsize \textcolor{red}{0.00 \tiny -0.54} & \scriptsize \textcolor{black}{0.98 \tiny -0.01} & \scriptsize \textcolor{red}{0.11 \tiny -0.81} \\ 
 & \tiny TL & \scriptsize \textcolor{red}{0.46 \tiny -0.52} & \scriptsize \textcolor{red}{0.64 \tiny -0.30} & \scriptsize \textcolor{red}{0.20 \tiny -0.39} & \scriptsize \textcolor{red}{0.00 \tiny -0.54} & \scriptsize \textcolor{black}{0.98 \tiny -0.01} & \scriptsize \textcolor{red}{0.17 \tiny -0.75} \\ 
 & \tiny X & \scriptsize \textcolor{red}{0.46 \tiny -0.52} & \scriptsize \textcolor{black}{0.93 \tiny -0.01} & \scriptsize \textcolor{black}{0.61 \tiny +0.02} & \scriptsize \textcolor{red}{0.00 \tiny -0.54} & \scriptsize \textcolor{black}{0.98 \tiny -0.01} & \scriptsize \textcolor{red}{0.31 \tiny -0.61} \\ \hline 
\SetCell[r=3]{r} {\footnotesize SG2 \\ \tiny{TPR}} & \tiny FB & \scriptsize \textcolor{red}{0.66 \tiny -0.33} & \scriptsize \textcolor{red}{0.85 \tiny -0.15} & \scriptsize \textcolor{red}{0.72 \tiny -0.28} & \scriptsize \textcolor{red}{0.00 \tiny -1.00} & \scriptsize \textcolor{red}{0.80 \tiny -0.14} & \scriptsize \textcolor{red}{0.02 \tiny -0.91} \\ 
 & \tiny TL & \scriptsize \textcolor{red}{0.48 \tiny -0.51} & \scriptsize \textcolor{red}{0.93 \tiny -0.07} & \scriptsize \textcolor{red}{0.69 \tiny -0.31} & \scriptsize \textcolor{red}{0.01 \tiny -0.99} & \scriptsize \textcolor{red}{0.88 \tiny -0.06} & \scriptsize \textcolor{red}{0.07 \tiny -0.86} \\ 
 & \tiny X & \scriptsize \textcolor{red}{0.48 \tiny -0.51} & \scriptsize \textcolor{black}{1.00 \tiny +0.00} & \scriptsize \textcolor{black}{0.98 \tiny -0.02} & \scriptsize \textcolor{red}{0.04 \tiny -0.96} & \scriptsize \textcolor{black}{0.91 \tiny -0.03} & \scriptsize \textcolor{red}{0.17 \tiny -0.76} \\ \hline 
\SetCell[r=3]{r} {\footnotesize SG3 \\ \tiny{TPR}} & \tiny FB & \scriptsize \textcolor{black}{0.61 \tiny +0.02} & \scriptsize \textcolor{blue}{0.46 \tiny +0.46} & \scriptsize \textcolor{blue}{0.78 \tiny +0.78} & \scriptsize \textcolor{black}{0.00 \tiny -0.02} & \scriptsize \textcolor{black}{0.96 \tiny +0.00} & \scriptsize \textcolor{red}{0.01 \tiny -0.81} \\ 
 & \tiny TL & \scriptsize \textcolor{red}{0.14 \tiny -0.45} & \scriptsize \textcolor{black}{0.00 \tiny +0.00} & \scriptsize \textcolor{blue}{0.25 \tiny +0.25} & \scriptsize \textcolor{black}{0.00 \tiny -0.02} & \scriptsize \textcolor{black}{0.93 \tiny -0.03} & \scriptsize \textcolor{red}{0.02 \tiny -0.80} \\ 
 & \tiny X & \scriptsize \textcolor{red}{0.08 \tiny -0.51} & \scriptsize \textcolor{black}{0.01 \tiny +0.01} & \scriptsize \textcolor{blue}{0.53 \tiny +0.53} & \scriptsize \textcolor{black}{0.00 \tiny -0.02} & \scriptsize \textcolor{black}{0.97 \tiny +0.01} & \scriptsize \textcolor{red}{0.06 \tiny -0.76} \\ \hline 
\SetCell[r=3]{r} {\footnotesize SD1.5 \\ \tiny{TPR}} & \tiny FB & \scriptsize \textcolor{red}{0.83 \tiny -0.08} & \scriptsize \textcolor{black}{0.97 \tiny -0.02} & \scriptsize \textcolor{black}{0.69 \tiny -0.03} & \scriptsize \textcolor{red}{0.08 \tiny -0.35} & \scriptsize \textcolor{black}{0.86 \tiny -0.02} & \scriptsize \textcolor{red}{0.77 \tiny -0.11} \\ 
 & \tiny TL & \scriptsize \textcolor{red}{0.84 \tiny -0.07} & \scriptsize \textcolor{black}{0.96 \tiny -0.03} & \scriptsize \textcolor{black}{0.74 \tiny +0.02} & \scriptsize \textcolor{red}{0.06 \tiny -0.37} & \scriptsize \textcolor{black}{0.90 \tiny +0.02} & \scriptsize \textcolor{red}{0.78 \tiny -0.10} \\ 
 & \tiny X & \scriptsize \textcolor{black}{0.91 \tiny +0.00} & \scriptsize \textcolor{black}{0.99 \tiny +0.00} & \scriptsize \textcolor{black}{0.71 \tiny -0.01} & \scriptsize \textcolor{black}{0.43 \tiny +0.00} & \scriptsize \textcolor{black}{0.87 \tiny -0.01} & \scriptsize \textcolor{black}{0.88 \tiny +0.00} \\ \hline 
\SetCell[r=3]{r} {\footnotesize SD2.1 \\ \tiny{TPR}} & \tiny FB & \scriptsize \textcolor{red}{0.67 \tiny -0.11} & \scriptsize \textcolor{red}{0.61 \tiny -0.17} & \scriptsize \textcolor{blue}{0.43 \tiny +0.06} & \scriptsize \textcolor{red}{0.11 \tiny -0.30} & \scriptsize \textcolor{red}{0.77 \tiny -0.08} & \scriptsize \textcolor{red}{0.80 \tiny -0.10} \\ 
 & \tiny TL & \scriptsize \textcolor{red}{0.66 \tiny -0.12} & \scriptsize \textcolor{red}{0.59 \tiny -0.19} & \scriptsize \textcolor{blue}{0.44 \tiny +0.07} & \scriptsize \textcolor{red}{0.11 \tiny -0.30} & \scriptsize \textcolor{black}{0.83 \tiny -0.02} & \scriptsize \textcolor{red}{0.81 \tiny -0.09} \\ 
 & \tiny X & \scriptsize \textcolor{black}{0.78 \tiny +0.00} & \scriptsize \textcolor{black}{0.78 \tiny +0.00} & \scriptsize \textcolor{black}{0.39 \tiny +0.02} & \scriptsize \textcolor{black}{0.42 \tiny +0.01} & \scriptsize \textcolor{black}{0.83 \tiny -0.02} & \scriptsize \textcolor{black}{0.90 \tiny +0.00} \\ \hline 
\SetCell[r=3]{r} {\footnotesize SD3 \\ \tiny{TPR}} & \tiny FB & \scriptsize \textcolor{red}{0.57 \tiny -0.06} & \scriptsize \textcolor{black}{0.67 \tiny +0.03} & \scriptsize \textcolor{blue}{0.80 \tiny +0.48} & \scriptsize \textcolor{red}{0.04 \tiny -0.51} & \scriptsize \textcolor{red}{0.76 \tiny -0.08} & \scriptsize \textcolor{red}{0.78 \tiny -0.09} \\ 
 & \tiny TL & \scriptsize \textcolor{red}{0.54 \tiny -0.09} & \scriptsize \textcolor{red}{0.55 \tiny -0.09} & \scriptsize \textcolor{blue}{0.74 \tiny +0.42} & \scriptsize \textcolor{red}{0.06 \tiny -0.49} & \scriptsize \textcolor{black}{0.82 \tiny -0.02} & \scriptsize \textcolor{red}{0.77 \tiny -0.10} \\ 
 & \tiny X & \scriptsize \textcolor{red}{0.38 \tiny -0.25} & \scriptsize \textcolor{red}{0.51 \tiny -0.13} & \scriptsize \textcolor{blue}{0.56 \tiny +0.24} & \scriptsize \textcolor{red}{0.12 \tiny -0.43} & \scriptsize \textcolor{black}{0.84 \tiny +0.00} & \scriptsize \textcolor{red}{0.77 \tiny -0.10} \\ \hline 
\SetCell[r=3]{r} {\footnotesize SDXL \\ \tiny{TPR}} & \tiny FB & \scriptsize \textcolor{black}{0.96 \tiny -0.03} & \scriptsize \textcolor{black}{1.00 \tiny +0.00} & \scriptsize \textcolor{black}{0.97 \tiny -0.02} & \scriptsize \textcolor{red}{0.21 \tiny -0.79} & \scriptsize \textcolor{black}{0.98 \tiny -0.02} & \scriptsize \textcolor{black}{0.97 \tiny -0.03} \\ 
 & \tiny TL & \scriptsize \textcolor{black}{0.96 \tiny -0.03} & \scriptsize \textcolor{black}{1.00 \tiny +0.00} & \scriptsize \textcolor{black}{0.98 \tiny -0.01} & \scriptsize \textcolor{red}{0.32 \tiny -0.68} & \scriptsize \textcolor{black}{1.00 \tiny +0.00} & \scriptsize \textcolor{black}{0.96 \tiny -0.04} \\ 
 & \tiny X & \scriptsize \textcolor{black}{0.95 \tiny -0.04} & \scriptsize \textcolor{black}{1.00 \tiny +0.00} & \scriptsize \textcolor{black}{0.98 \tiny -0.01} & \scriptsize \textcolor{red}{0.62 \tiny -0.38} & \scriptsize \textcolor{black}{1.00 \tiny +0.00} & \scriptsize \textcolor{black}{0.98 \tiny -0.02} \\ \hline 
\SetCell[r=3]{r} {\footnotesize FLUX.1 \\ \tiny{TPR}} & \tiny FB & \scriptsize \textcolor{red}{0.81 \tiny -0.11} & \scriptsize \textcolor{black}{0.94 \tiny +0.00} & \scriptsize \textcolor{blue}{0.93 \tiny +0.07} & \scriptsize \textcolor{red}{0.03 \tiny -0.83} & \scriptsize \textcolor{red}{0.91 \tiny -0.07} & \scriptsize \textcolor{red}{0.82 \tiny -0.11} \\ 
 & \tiny TL & \scriptsize \textcolor{red}{0.85 \tiny -0.07} & \scriptsize \textcolor{black}{0.92 \tiny -0.02} & \scriptsize \textcolor{blue}{0.94 \tiny +0.08} & \scriptsize \textcolor{red}{0.05 \tiny -0.81} & \scriptsize \textcolor{black}{0.98 \tiny +0.00} & \scriptsize \textcolor{red}{0.80 \tiny -0.13} \\ 
 & \tiny X & \scriptsize \textcolor{red}{0.80 \tiny -0.12} & \scriptsize \textcolor{black}{0.94 \tiny +0.00} & \scriptsize \textcolor{blue}{0.95 \tiny +0.09} & \scriptsize \textcolor{red}{0.18 \tiny -0.68} & \scriptsize \textcolor{black}{0.98 \tiny +0.00} & \scriptsize \textcolor{red}{0.81 \tiny -0.12} \\ \hline \hline 
\end{tblr}
\label{tab:results_shared}
\end{table}

In Table \ref{tab:results_shared}, we report the {TPR and TNR} achieved when training on the same non-shared images from Sect.~\ref{sec:non-shared} and testing on images that have been shared on social networks.

Starting with the CNN-based detectors, we observe that NPR's fake image detection capabilities are severely hindered by all social networks image processing. This is presumably due to NPR’s reliance on ephemeral low-level traces, which are heavily affected by image compression (not considered in the original paper) and resizing. {The only exceptions are images shared on X that were generated by SD1.5 and SD2.1, which are not modified by X, as discussed in Sect.~\ref{sec:dataset_shared}.}

Regarding the other CNN-based methods, while MISLNet exhibits a substantial increase in {TPR} for fake images, its {TNR} for real images decreases significantly. This suggests a shift in the feature distribution used for detection. 

R50-ND, while affected by social network image sharing, thanks to its extensive data augmentation shows a better resilience compared with other CNN-based detectors. Conversely, while {R50-E2P} generally outperforms MISLNet and NPR, its performance on GAN-generated and FFHQ real images are worst compared to R50-ND. This last result, as confirmed by the performance of R50-ND on SG3 images shared via Facebook, can be attributed to the refined low-level feature representation learned by R50-ND during its re-training.

Examining the results of the ViT-based methods, P2G performs significantly worse on shared data compared to CLIP-D and most CNN-based detectors. This is particularly evident for GAN-generated images and is likely due to the absence of any JPEG data augmentation in P2G’s training pipeline.

In contrast, CLIP-D, which incorporates JPEG data augmentation, achieves the best performance on shared images. However, especially for images shared on Facebook, the problem remains far from solved as margins for improvement still exist, and {TPR and TNR} losses for FFHQ real images, SG2, SD2.1, SD3, and FLUX.1 approach or exceed $10\%$.

To conclude, Table~\ref{tab:results_shared} highlights that all social networks have a significant impact on the selected fake image detectors, particularly affecting those that rely on pixel-level dependencies (e.g., NPR) and those that do not employ JPEG compression or resizing as part of their training data augmentation (e.g., P2G). Furthermore, Facebook appears to be the most disruptive social network, with {true positive and negative rates} losses ranging from $10\%$ to $100\%$ across all detectors. As indicated in Table \ref{tab:dataset_social}, and discussed in Sect.~\ref{sec:dataset_shared}, this is likely due to Facebook’s implementation of a content-aware image processing scheme.

\section{Conclusions}
\label{sec:conclusions}

In this paper, we introduce TrueFake, a dataset of 600,000 AI-generated {and real} images, including 180,000 shared on three social networks (i.e., Facebook, X, and Telegram) to assess the impact of social media processing on the performance of SoA fake image detectors.

We evaluate five recent and accurate detectors by comparing their {TPR and TNR} in controlled conditions (images not shared on social networks) versus real-world scenarios (images shared online). While most detectors generalize well in ideal settings, all experience {TPR and TNR} degradation in real-world conditions. Methods relying on pixel-level inconsistencies (e.g., NPR), lacking pre-training on large real-image datasets (e.g., MISLNet), or omitting JPEG and resizing augmentations (e.g., P2G) are particularly affected. In contrast, CLIP-D, which integrates all these strategies, consistently outperforms others across both settings. Nevertheless, while CLIP-D achieves a reasonable baseline performance on shared images, for 50\% of the real and fake image classes studied in this paper, its {TPR and TNR} exhibits losses approaching or exceeding $10\%$ on at least one social network.

Finally, while preliminary, the results obtained {by the here proposed} R50-E2P suggest a new direction for fake image detection. Specifically, they indicate that any network pre-trained on a sufficiently large dataset of real images can be used as a feature encoder, without restricting this role to Vision Transformers used in Large Language Models, as was originally the case for CLIP-D.

\section*{Acknowledgment}
{This work was partially supported by the European Union under the Italian National Recovery and Resilience Plan (NRRP) of NextGenerationEU (PE00000014 - program “SERICS”) and project Deepfake Detection 2.0 funded by Fondazione VRT. }

\bibliographystyle{ieeetr}
\bibliography{main}
\end{document}